# Surfactant concentration modulates the motion and placement of microparticles in an inhomogeneous electric field

Marcos Masukawa[a], Masayuki Hayakawa[b,c] and Masahiro Takinoue[*a,b]

[a]Department of Computer Science, Tokyo Institute of Technology, 4259 Nagatsuta-cho, Midori-ku, Yokohama, Kanagawa, 226-8502, Japan.

[b]Department of Computational Intelligence and Systems Science, School of Computing, Tokyo Institute of Technology, 4259 Nagatsuta-cho, Midori-ku, Yokohama, Kanagawa226-8502, Japan.

[c]RIKEN Center for Biosystems Dynamics Research, Kobe, Hyogo 650-0047, Japan.

*Correspondance to M.T. E-mail: takinoue@c.titech.ac.jp.

**Abstract**

This study examined the effects of surfactants on the motion and positioning of microparticles in an inhomogeneous electric field. The microparticles were suspended in oil with a surfactant and the electric field was generated using sawtooth-patterned electrodes. The microparticles were trapped, oscillated, or attached to the electrodes. The proportion of microparticles in each state was defined by the concentration of surfactant and the voltage applied to the electrodes. Based on the trajectory of the microparticles in the electric field, a newly developed physical model in which the surfactant was adsorbed on the microparticles allowed the microparticles to be charged by contact with the electrodes, with either positive or negative charges, while the non-adsorbed surfactant micellizing in the oil contributed to charge relaxation. A simulation based on this model showed that the charging and charge relaxation, as modulated by the surfactant concentration, can explain the trajectories and proportion of the trapped, oscillating, and attached microparticles. These results will be useful for the development of novel self-assembly and transport technologies and colloids sensitive to electricity.

**Introduction**

Techniques for manipulating microparticles are important in physical, chemical, and biological research[1]. Fundamentally, the ability to control small particles in small volumes can help elucidate the mechanisms that operate at the µm scale. From a practical point of view, these mechanisms can be explored to build sensors and actuators—thereby extending the capabilities of microfluidic devices and display technologies and bridging the macro- and nanoscale[2].

Researchers have used optical tweezers[3], surface acoustic waves[4], chemical gradients[5], and magnetic[6] and electric fields[7–14], among other non-contact methods, to manipulate microparticles. Electric fields represent an attractive method for controlling microparticles in particular. For instance, electrodes can be designed to produce specific electric fields that can be quickly modulated via changes in the frequency and amplitude of the applied voltage[15]. Microparticle movement in an electric field is often referred to as electrokinetics, and distinct mechanisms govern the interactions of microparticles with electric fields for displacement in a controlled manner[16,17]. Examples include dielectrophoresis via the application of an inhomogeneous electric field[11,15,18,19] and contact charge electrophoresis, when the particle charge is modified by contact with a charged object[9,10,20].

To control a particle using an electric field, easily modified electric and dielectric properties of the particle and surrounding medium are desirable. Surfactants can modify the medium conductivity[21–24] and charge microparticles suspended in apolar liquids[24–32]. Although not fully understood, the modification of electrical properties of apolar colloids by surfactant addition is of great importance in industry as it is used to control the electrical properties of ink in printing processes[21]. This is also used to prepare electrophoretic ink displays to control the position of the pigment microparticles[8] and to set their position in an organized manner to form colloidal crystals[33], which can be used as photonic materials. Therefore, significant interest has been generated in the chemical synthesis of novel surfactants suitable for particle control[27], to clarify the role of surfactants in electrokinetics and extend its applications. However, the effect of surfactant concentration on the electrokinetics of microparticles remains unclear due to the challenges of producing a model that considers the surfactant concentration effect on both contact charge electrophoresis and charge relaxation. Such a model would be useful to predict the ideal surfactant concentration for the manipulation of microparticles by electric fields.

Herein, the effect of concentration of a neutral surfactant on microparticles suspended in an apolar liquid subject to an inhomogeneous electric field was studied. Based on the experimental observations, a model of particle charging and charge relaxation modulated by the surfactant for definition of displacement and position of the microparticles was developed and investigated.

**Materials and methods**

**Sample preparation of microparticle suspension in oil with a surfactant**

Liquid paraffin (Wako, 128-04375) was used as an oil phase, Sorbitan Monooleate (Span 80) (TCI chemicals, S0060) as a neutral surfactant, and polystyrene microbeads 20 µm in diameter (Micromod, 01-00-204) as microparticles. A stock solution was prepared by vortexing liquid paraffin and 0.0001% (w/w) Span 80, sonicating the mixture at 45°C for 1 h and adding microparticles. A low concentration of microparticles, 0.01% (w/w), was used to minimize the interaction between microparticles. The stock was vortexed and sonicated again at 45°C for 1 h. Afterwards, microparticle suspensions with different concentrations of surfactant were prepared by adding Span 80 to the stock solution, vortexing, sonicating at 45°C for 1 h, and leaving the mixture to equilibrate for 1 h. Suspensions of microparticles were prepared in liquid paraffin with Span 80 concentrations ranging from 0.0001 to 5% (w/w).

**Sample preparation of the fluorescent reverse micelles**

Reverse micelles of Span 80 aggregates containing fluorescein were prepared following the protocol described by Anton et al. (2011)[34]. First, a saturating amount of fluorescein (Sigma-Aldrich, F6377-100G) was added to liquid paraffin containing 0.1% (w/w) Span 80. The dispersion was then vortexed and sonicated at 45°C. To remove fluorescein crystals, the dispersion was centrifuged for 10 min at 2000×g and the supernatant was collected, with the process being repeated three times. Afterwards, the fluorescent dispersion of Span 80 was diluted to 0.005% (w/w) and the microparticles were added. The samples were subsequently vortexed and sonicated at 45°C for 1 h. To prepare the fluorescent samples with different surfactant concentrations, Span 80 was added without fluorescein to the desired concentration. Fluorescent samples in concentrations of 0.005% (w/w) (no additional surfactant), 0.05% (w/w), 0.5% (w/w), and 5% (w/w) were prepared.

**Microelectrode fabrication**

Interdigitated microelectrodes with sawtooth edges and a 70 µm gap between the teeth (Fig. 1A,B) were prepared using the lift-off method[35]. Briefly, a S1818G photoresist (Microchem) was spin coated onto a micro cover glass No. 5 (Matsunami) treated with plasma oxidation (90 s on an ion bombarder, Vacuum Device Co., Ltd) and silylated with hexamethyldisilazane (HMDS, Wako AWK3814) via vapour phase deposition[36]. The photoresist was spin coated at a maximum spin frequency of 3000 rpm for 30 s (Opticoat SpinCoater, Mikasa). Subsequently, the slide was pre-baked for 1 min at 115°C, cooled to room temperature, and exposed using a maskless pattern generator with resolution of 3 µm (µPG 101, Heidelberg Instruments; laser wavelength, 375 nm). Afterwards, the photoresist was developed with tetramethylammonium hydroxide solution 2.38% (OFPR-NMD-3, Tokyo Ohka Kogyo Co., Ltd.) and cleaned with isopropyl alcohol (IPA, Kanto Chemical Co. Inc. JIS K8839). The developed slide was then coated sequentially with chromium and gold using a metal evaporator (VE2012 TMP vacuum evaporator, Vacuum Device Co., Ltd). Finally, the undeveloped photoresist was removed with acetone (Wako, DSG4138), revealing a sawtooth pattern. Supplementary Method 1 contains additional details regarding the sample preparation.

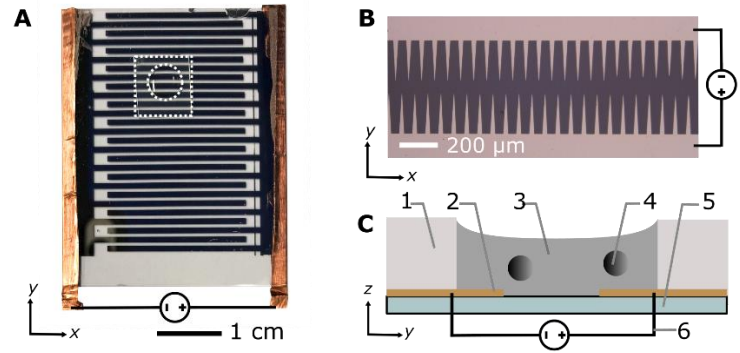

**Figure 1** – Experimental set-up. (A) Photograph of the device from the top, dashed lines indicate the gasket used to contain the sample. (B) Photograph of electrodes from the top against a dark background. (C) Illustration of side view. 1. PDMS gasket; 2. Electrode with sawtooth edge; 3. Liquid paraffin with Span 80; 4. Polystyrene microparticle; 5. Glass slide; 6. Direct current voltage.

**Microscopic measurements**

To track the microparticle position, 12 µL samples were placed directly on top of the microelectrodes and placed in a polydimethylsiloxane (PDMS) gasket with a 5 mm diameter (Fig. 1A,C). Direct current (DC) voltages of 100, 200, and 300 V were sequentially applied to the electrodes using a DC power supply (GWInstek, GPR30H10D). The measurements were performed in triplicate for all voltages and surfactant concentrations and, on average, the trajectory of 51±21 microparticles per surfactant concentration per voltage were analysed for each experiment. The position of the microparticles was recorded using a phase contrast microscope (Olympus, CKX41) and digital camera (Canon, EOS60D). From the recorded videos, the positions of the individual microparticles were tracked using custom image analysis software and the Python package Trackpy[37] and the motion patterns of the beads were classified (Supplementary Method 2). Fluorescent samples were used to observe the location of the Span 80 aggregates and measure the ratio of the dispersed to adsorbed surfactant on the microparticles (Supplementary Method 3). A fluorescence microscope (Olympus, IX71) equipped with a sCMOS camera (Andor, Zyla) was used to observe the fluorescent samples. The microscope was equipped with a mercury lamp source, mirror unit with 470–490 nm band-pass excitation filter, 505 nm dichroic mirror, and 510–550 nm band-pass emission filter (Olympus, NIBA).

**Numerical simulations**

The finite element software COMSOL Multiphysics (COMSOL Inc., v4.3), was used to calculate the electric and dielectric fields generated by the saw tooth electrodes at 200 V. Using the fields, the trajectory of 500 microparticles with random initial conditions under 12 different surfactant concentrations were monitored, totalling 6000 simulated trajectories. For the simulation, a custom Python script with the Runge-Kutta $4^{th}$ order was used for numerical integration. After simulation, the robustness of the obtained results was determined by comparing 10 random subsets of trajectories. The theory for the simulation is discussed in the Results and Supplementary Discussions 1 and 2, while the choice of parameters is discussed in the Supplementary Note 1.

**Results and discussion**

When an electric field was applied to the dispersed microparticles in oil with surfactant using sawtooth electrodes, the microparticles exhibited one of three motion patterns regarding their position and displacement as follows: 'trapped', when the microparticle remained between the electrodes without touching them Fig. 2A; 'oscillating', when the microparticles moved periodically between the electrodes Fig. 2B; and 'attached', when the microparticles remained in close contact with one of the electrodes Fig. 2C. The microparticles in the same sample did not all exist in the same state and, on occasion, a microparticle would switch from one state to another. However, the proportion of microparticles in a certain state within a sample remained approximately constant depending on the surfactant concentration and applied voltage (Fig. 3). The oscillating microparticles were most abundant when the surfactant concentration was approximately 0.05% (w/w), while most microparticles were in the attached state when the surfactant concentration was <0.005% (w/w) or >0.5% (w/w).

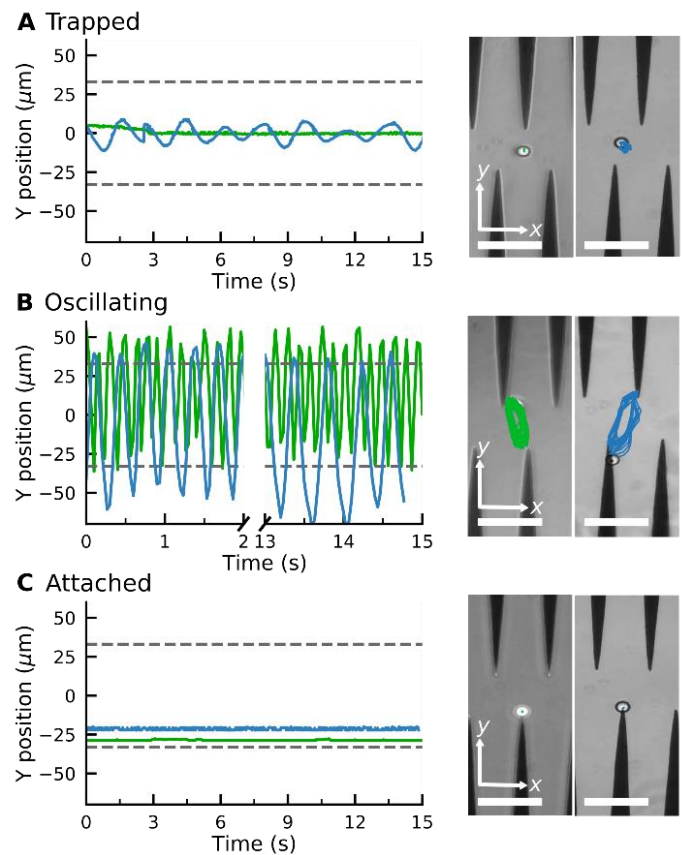

**Figure 2** – Microparticle classification into three patterns of motion based on microparticle trajectory (left) obtained from interference microscopy (right). The microparticles were (A) trapped between the electrodes, (B) oscillating between the electrode tips, or (C) attached to the electrode edge. The trajectory of sample microparticle 1 (━) and sample microparticle 2 (━). Scale bar = 100 µm.

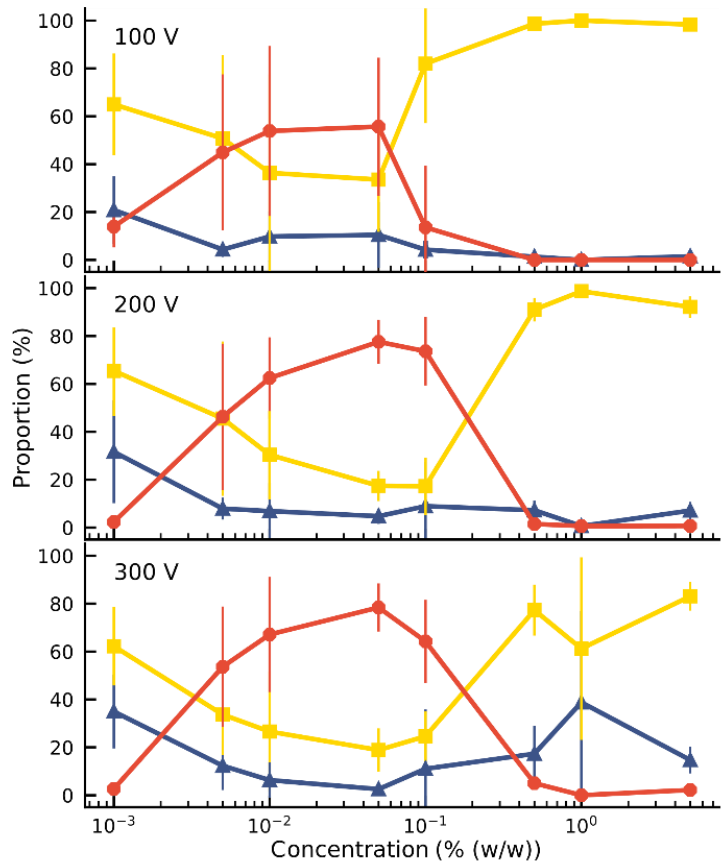

**Figure 3** – Proportion of trapped (▲), oscillating (●), and attached (■) microparticles in a sample according to the surfactant concentration and applied voltage. The solid lines are guides to the eye. Error bars indicate standard deviation.

To investigate the role of the surfactant in the mechanism of motion, reverse micelles of Span 80 with a loaded fluorescent dye were used to observe the location of the surfactant within the dispersion. The surfactant reverse micelles with fluorescein were initially dispersed homogeneously, but they subsequently adsorbed on the microparticle surfaces (Fig. 4A). The adsorption was not homogeneous among the microparticles and varied depending on the final concentration of surfactant (Fig. 4A). Fluorescein sodium is a salt and does not dissolve in oil or adsorb directly on the surface of the polystyrene microparticles; therefore, the fluorescence density was assumed to be proportional to the local concentration of surfactant.

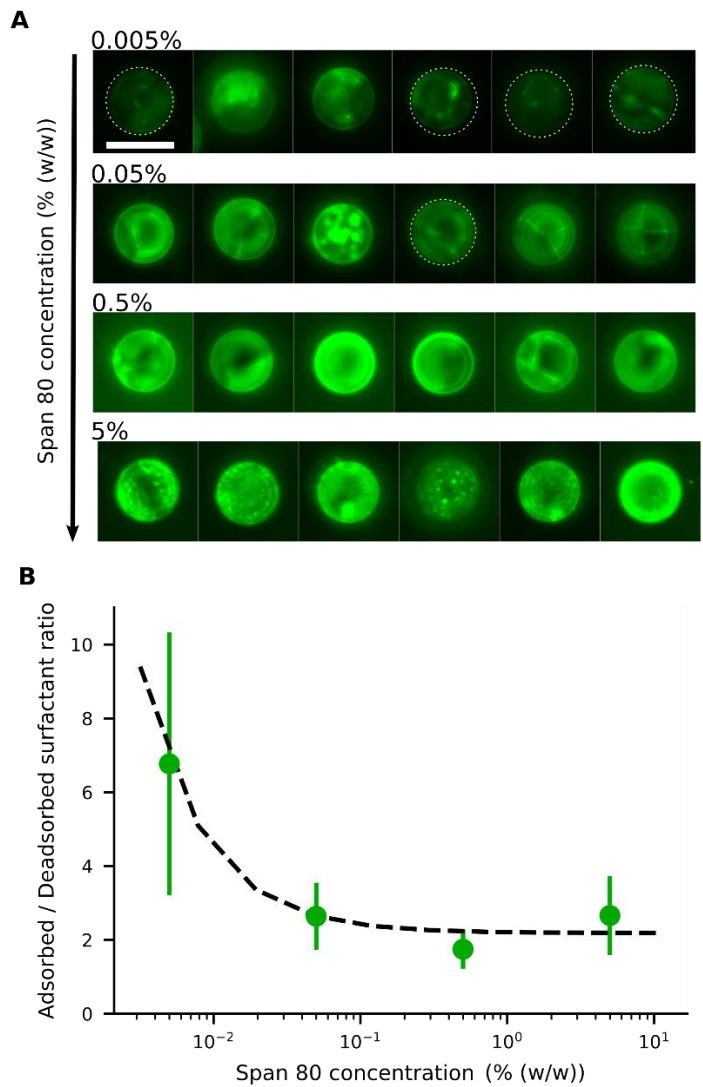

**Figure 4** – Adsorption of reverse micelles loaded with fluorescein on the microparticles. (A) Fluorescence images of microparticles with different concentrations of Span 80; images in the same row were taken from the same sample. Dotted lines delimit the contour of microparticles with low fluorescence. Scale bar = 20 µm. (B) Ratio of surfactant adsorbed on the microparticles and dispersed in the medium according to surfactant concentration as measured by fluorescence (●) and according to the Langmuir adsorption model, fitted to the experimental points (▬ dashed line).

By measuring the fluorescence ratio between the microparticles and oil, the ratio of the surfactant on the microparticle surface to that in the oil was determined (Fig. 4B). At low surfactant concentrations, the ratio of adsorbed surfactant on the desorbed surfactant was high, but the surface of the microparticles became saturated with a constant ratio as the surfactant concentration increased. The ratio of absorbed to desorbed surfactant, derived from the Langmuir adsorption model[38], is given by Eq. (1):

$$\frac{C_{ad}}{C_d} = \frac{C_{max}K_{ad}}{1 + K_{ad}(C_t - C_{max})} \quad (1)$$

where $C_{ad}$ is the concentration of adsorbed surfactant on the microparticle surface; $C_d$ is the surfactant concentration dissolved in the medium; $C_{max}$ is the maximum possible concentration of adsorbed surfactant; $K_{ad}$ is the adsorption equilibrium constant; and $C_t$ is the total surfactant concentration (see Supplementary Discussion 1). To fit the experimental data, a baseline was used (Fig. 4B).

As the reverse micelles can stabilize charges in apolar media[22,25,30,39,40], two properties of the suspension depend on surfactant concentration: the microparticle charge and medium conductivity. These properties depend mainly on the local surfactant concentration; that is, the microparticle charge is limited by the amount of surfactant adsorbed, $C_{ad}$, while the medium conductivity, $\sigma$, is limited by the amount of surfactant in the medium, $C_d$. The charge ceiling of the microparticle was defined as the charge carrying capacity $Q(C_t)$, as in Eq. (2) below:

$$Q(C_t) = q_{ar} \frac{K_{ad}C_t}{1 + K_{ad}C_t} \quad (2)$$

where $q_{ar}$ is the charge per area per adsorbed surfactant unit (see Fig. 5 and Supplementary Discussion 1).

The conductivity of a medium is proportional to the surfactant concentration in it (observed for Span 80 in hexane[28] and isopar-L[40]), which can be defined in terms of the charge relaxation law: if a charged particle is suspended in a fluid with uniform conductivity, $\sigma$, and dielectric constant, $\varepsilon_m$, its charge decays with a time constant $\tau_d = \varepsilon_m/\sigma$[41]. Therefore, considering the surfactant effects on the medium conductivity, Eq. (3) can be obtained:

$$\tau_d(C_t) = \frac{\epsilon_m}{\sigma_0 + q_s C_t} \qquad (3)$$

where $\sigma_0$ is the conductivity of pure liquid paraffin, $q_s$ is the rate of conductivity increase per surfactant unit in the medium, and $C_d \approx C_t$ for low microparticle concentrations (Fig. 5).

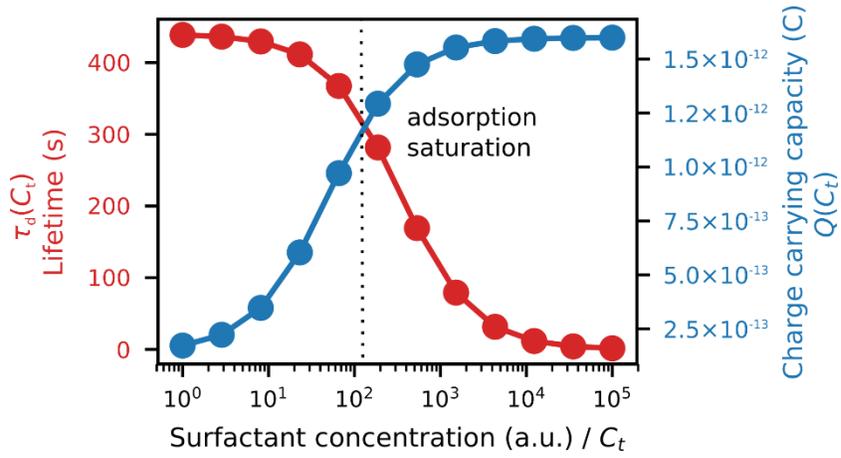

**Figure 5** – Model on how surfactant concentration affects the parameters used in the simulation of microparticle trajectory: charge carrying capacity (●) and charge relaxation lifetime (●). Parameters used in the simulation: $q_{ar} \approx 1.48 \times 10^{-12}$; $K_{ad} \approx 2.74 \times 10^{-2}$; $\sigma_0 \approx 8.88 \times 10^{-3}$; $q_s \approx 4.44 \times 10^{-5}$.

From the experimental trajectories of the microparticles, as those shown in Fig. 2, it is possible to guess the sign and magnitude of the microparticle charge (Fig. 6A). For example, a microparticle attaches to the electrodes when its instant charge $q$ is smaller than a charge threshold $q_1$, $q_1$ such that the electric force is smaller than the dielectric and viscous forces. If the charge is above this threshold, the particles migrate to the electrodes of opposite charge. However, when the particle is migrating, it can eventually be trapped between the electrodes, where the dielectric force is null, if the electric force is smaller than the viscous force, that is, if the microparticle instant charge is smaller than $q_2$, $q_2$ being smaller than $q_1$. The oscillating microparticles show the microparticle can exchange charge when they touch the electrodes, while trapped microparticles show they can lose charge when they are not touching the electrode, as defined by the charge relaxation law (Fig. 6B). Eq. (4) was used to describe the change in charge after a microparticle touches an electrode, which is derived from the charging of a sphere by a unipolar current[41–45] (see Supplementary Discussion 2):

$$q(t') = \frac{t'/\tau_c(Q - q_a) + q_a}{1 + t'/\tau_c(1 - q_a/Q)} \quad (4)$$

where $t'$ is the time after a particle attaches to the electrode; $q(t')$ is the microparticle charge at time $t'$; $\tau_c$ is a charging rate parameter; $and$ $q_a$ is the charge of the particle when the particle attaches to the electrode. In Eq. (4), the charge $q(t')$ approaches $Q$ as $t' \to \infty$ and the rate of charging decreases as the charge approaches this limit. When the microparticles are not contacting the electrodes, their charge decreases exponentially, as described in Eq. (5):

$$q(t'') = q_d e^{-t''/\tau_d} \quad (5)$$

where $t''$ is the time after a particle detaches from the electrode; $\tau_d$ is the charge relaxation lifetime; and $q_d$ is the charge of the particle when the particle detaches from the electrode.

In our model, $Q$ and $\tau_d$ are functions of the surfactant concentration (Eqs. (2) and (3)). In that sense, the surfactant in an apolar medium promotes opposing mechanisms of particle charging and charge relaxation, both of which depend on concentration.

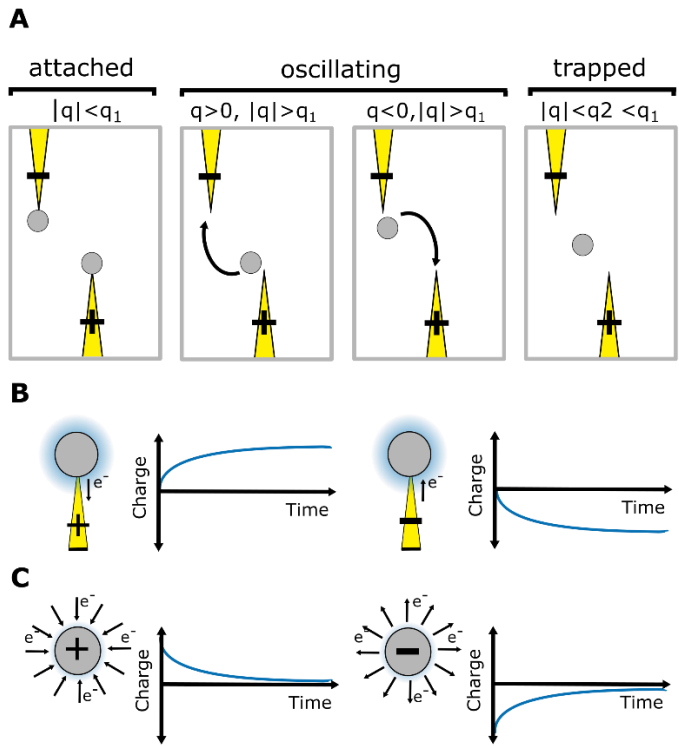

**Figure 6** – (A) Microparticle trajectory indicates the sign and magnitude of charge; $q$ is the particle instant charge, $q_1$ and $q_2$ are charge thresholds. (B) Microparticle charging. (C) Microparticle charge relaxation.

Using Eqs. (2)–(5) and parameters estimated from the literature (see Supplementary Note 1), the trajectories of 500 microparticles were simulated while varying the charge carrying capacity and charge relaxation lifetime, while emulating increasing surfactant concentrations (Fig. 7). To simulate the particle trajectory, the electric and dielectric fields were simulated using finite elements software. The electric force $\vec{F}_{el}$,

$$\vec{F}_{el}(\vec{x}, t) = q(t)\vec{E}(\vec{x}) \tag{6}$$

depends on the particle charge $q(t)$, where $t$ is time and $\vec{x}$ is the microparticle position. The dielectric force $\vec{F}_{dl}$, which originates from the interaction of the inhomogeneous electric field and particle dipole, can be given as follows:

$$\vec{F}_{dl}(\vec{x}) = 2\pi R^3 \epsilon_m \frac{\epsilon_p - \epsilon_m}{\epsilon_p + 2\epsilon_m} \nabla E^2(\vec{x}) \tag{7}$$

It is proportional to $R^3$, where $R$ is the particle radius and depends on $\varepsilon_m$ and $\varepsilon_p$, the dielectric constants of the medium and particle, respectively. In our system, $\epsilon_m > \epsilon_p$, which means the dielectric force attracts the microparticles towards the electrode tips, where the electric field is more divergent. Thereafter, the overdamped equation of motion can be constructed as follows:

$$\frac{d\vec{x}}{dt} 6\pi\eta R = \vec{F}_{el}(\vec{x}, t) + \vec{F}_{dl}(\vec{x}) \tag{8}$$

where the left side expresses the viscous force (Stokes drag force) predominant in microenvironments[9].

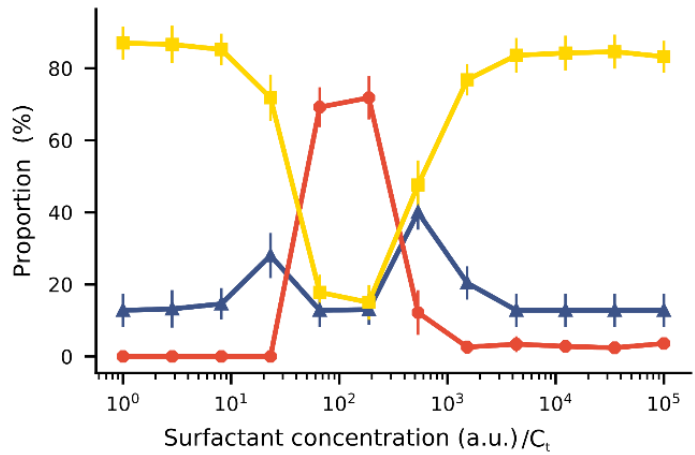

**Figure 7** – Simulated distribution of trapped (▲), oscillating (●), and attached (■) microparticles according to the surfactant concentration. Each point represents the average of 10 sets of 50 trajectories each. Error bars indicate standard deviation.

During the simulation, small variations in the initial conditions—such as initial position, velocity, and charge—yielded different trajectories (see Supplementary Note 2). To account for this observation, the simulations were performed using 500 initial conditions that were later randomly split into 10 sets; the results are shown in Fig. 7. The simulation suggests that the trajectory of a microparticle depends on its initial conditions, although the proportion of each motion pattern is approximately constant and defined by surfactant concentration.

The relationship between particle trajectory and charge was examined to better understand the origin of the different motion patterns. A representative simulation of a particle trajectory is shown in Fig. 8, with the model of local concentration of surfactant aggregates in the insets. The surfactant concentration changes the microparticle charge carrying capacity and relaxation lifetime, which changes the balance between the dielectric force (field shown in Fig. 8) and electric force. The dielectric force pushes the particle towards the electrode tips whereas the electric force depends on the position and particle charge, which changes dynamically due to charging and charge relaxation mechanisms enabled by the surfactant. When a microparticle is charged, it migrates towards the electrode with the opposite charge. Generally, attached microparticles were observed when the dielectric force was dominant and forced them close to the electrode tips. This occurred at very low and very high surfactant concentrations. At very low concentrations, the surfactant concentration is low both on the microparticle and in the medium and the microparticle charge is low. At very high surfactant concentrations, charge relaxation is rapid due to the increased amount of surfactant in the medium and microparticles have a low average charge. At marginally low and marginally high surfactant concentrations, trapped microparticles were more common. Trapped microparticles were observed when the charge decayed while the microparticles were close to the stable point between the electrodes, where the dielectric force is null. Oscillating microparticles were most common at intermediate surfactant concentrations, when the charge carrying capacity and charge relaxation lifetime were high. At this concentration range, the microparticle has sufficient charge to reach the opposite electrode without becoming trapped at the stable point. When the particle reaches the opposite electrode, it is charged with the opposite charge and thus the cycle is restarted.

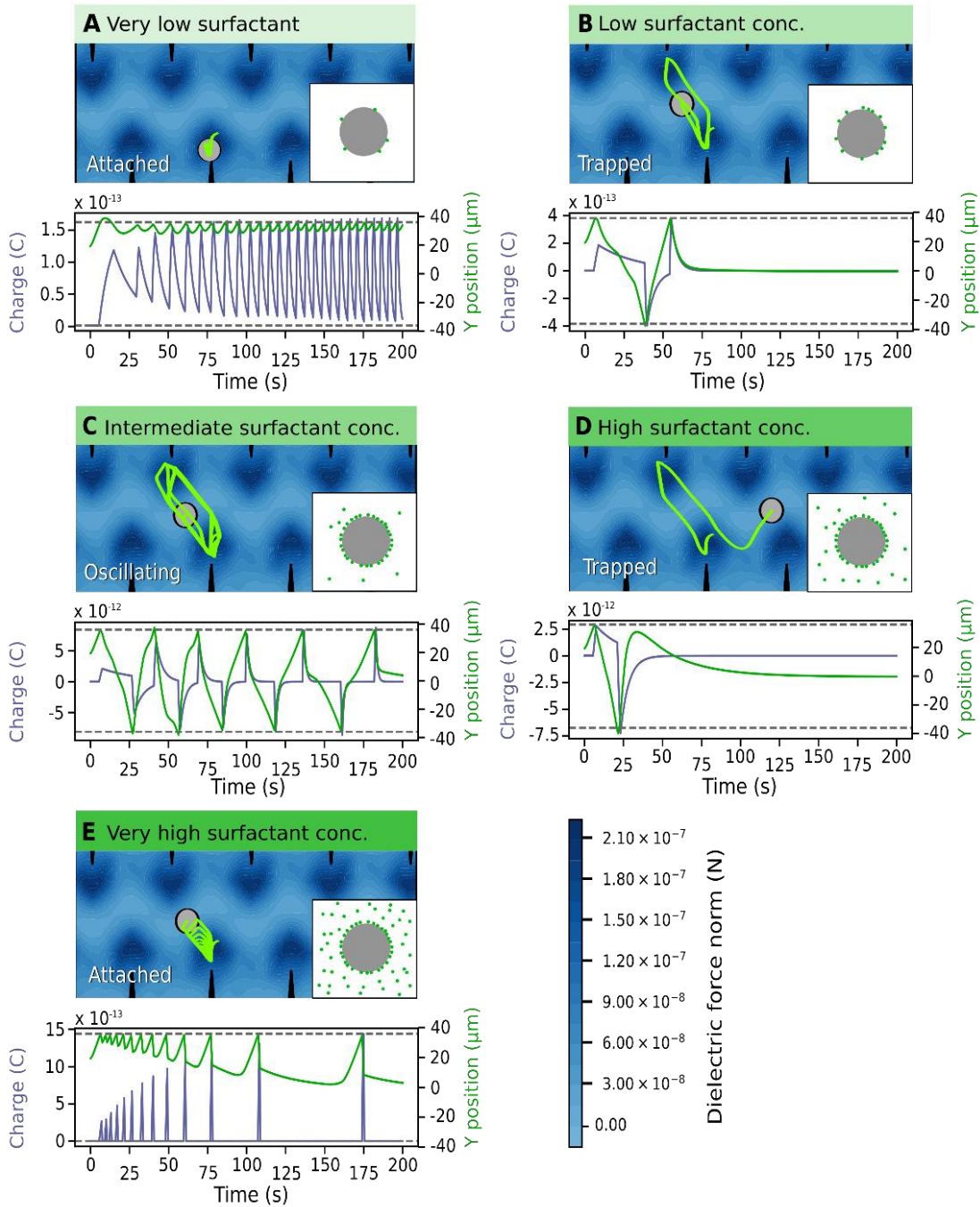

**Figure 8** – Simulated trajectory of a microparticle at various surfactant concentrations demonstrating the effect of changing the parameters charge carrying capacity (*Q)* and charge relaxation lifetime ($\tau_d$). The underplot shows the microparticle position (—) and charge (—) as a function of time in relation to the tip of the electrodes (--). (A) $Q \approx 1.75e^{-13}$ C, $\tau_d \approx$ 437.81 s; (B) $Q \approx 8.44\ e^{-13}$ C, $\tau_d \approx 357.11$ s ; (C) $Q \approx 1.26\ e^{-12}$ C, $\tau_d \approx 239.90$ s; (D) $Q \approx 1.56e^{-12}$ C, $\tau_d \approx 37.37$ s; (E) $Q \approx 1.59e^{-12}$ C, $\tau_d \approx 0.87$ s.

A comparison of Figs. 7 and 3 demonstrates the newly developed model succeeds in explaining the enhanced oscillation or trapping at certain surfactant concentration and the concentration thresholds. It should be cautioned that the model is two-dimensional, estimating forces on the microparticles that are not equivalent to what would be expected in actual experiments. For instance, in Fig. 8 the oscillating frequencies are significantly lower than those observed experimentally (Fig. 2). Furthermore, changes in other parameters of the system were not considered, but are expected upon increasing surfactant addition, such as the dielectric constant[46] and viscosity[47].

**Conclusions**

Surfactants change the electrokinetics of microparticles by enabling their charging and discharging in an electric field. Varying surfactant concentration changes the ratio of surfactant adsorbed on the microparticle to that in the medium, which charges the microparticles when in contact with electrodes and discharges the microparticle in the medium. The change in charge modifies the balance between the electric, dielectric, and viscous forces, creating a dynamic system wherein the microparticles display different motion patterns depending on surfactant concentration. This indicates that a static, inhomogeneous electric field is a versatile tool for particle control and surfactants can act as a mediator to integrate micromachines and digital devices. The further understating of microparticle charging mediated by surfactants may allow for the improved self-assembly of microstructures[48], motion of microrobots[49], and individual microparticle control[50] by designing novel electrode geometries. The newly developed model can also provide a rationale for the syntheses of novel surfactants for use in electric sensitive colloids[51] by tuning surfactant adsorption and charged aggregate formation.

**Acknowledgements**

This research was supported by JSPS KAKENHI grants to M.T. (No. JP17H01813, JP18K19834), Research Encouragement Grants from The Asahi Glass Foundation to M.T., Support for Tokyo Tech Advanced Researchers (STAR) to M.T., and Japanese Government (MEXT) Scholarship for foreign students to M.K.M.

Supplementary Materials for

**Surfactant concentration modulates the motion and placement of microparticles in an inhomogeneous electric field**

Marcos Masukawa[a], Masayuki Hayakawa[b,c] and Masahiro Takinoue[*ab]

[a]Department of Computer Science, Tokyo Institute of Technology, 4259 Nagatsuta-cho, Midori-ku, Yokohama, Kanagawa, 226-8502, Japan.

[b]Department of Computational Intelligence and Systems Science, School of Computing, Tokyo Institute of Technology, 4259 Nagatsuta-cho, Midori-ku, Yokohama, Kanagawa226-8502, Japan.

[c]RIKEN Center for Biosystems Dynamics Research, Kobe, Hyogo 650-0047, Japan.

*Correspondance to M.T. E-mail: takinoue@c.titech.ac.jp.

**This PDF file includes:**

    Supplementary Methods

    Figs. S1 to S3

    Supplementary Discussion

    Figs. S4 to S5

    Supplementary Notes

    Fig. S6

**Other Supplementary Materials for this manuscript includes the following:**

    Movies S1 to S6

**Supplementary Methods**

1) <u>Device design and experiment assembly</u>

The electrodes were designed using the software Rhinoceros (Robert McNeel & Associates, v4.0) and consisted of interdigitated electrodes with a saw-tooth edge, as shown in Fig. S1. This arrangement was used to maximize the number of electrodes on the slide. Copper tape and conductive adhesive were used to connect the electrodes to a direct current source. Twelve microliters of the sample were pipetted in the well, bubbles were removed with a blower, and the microparticles were observed from bottom to top of the slide using an inverted microscope.

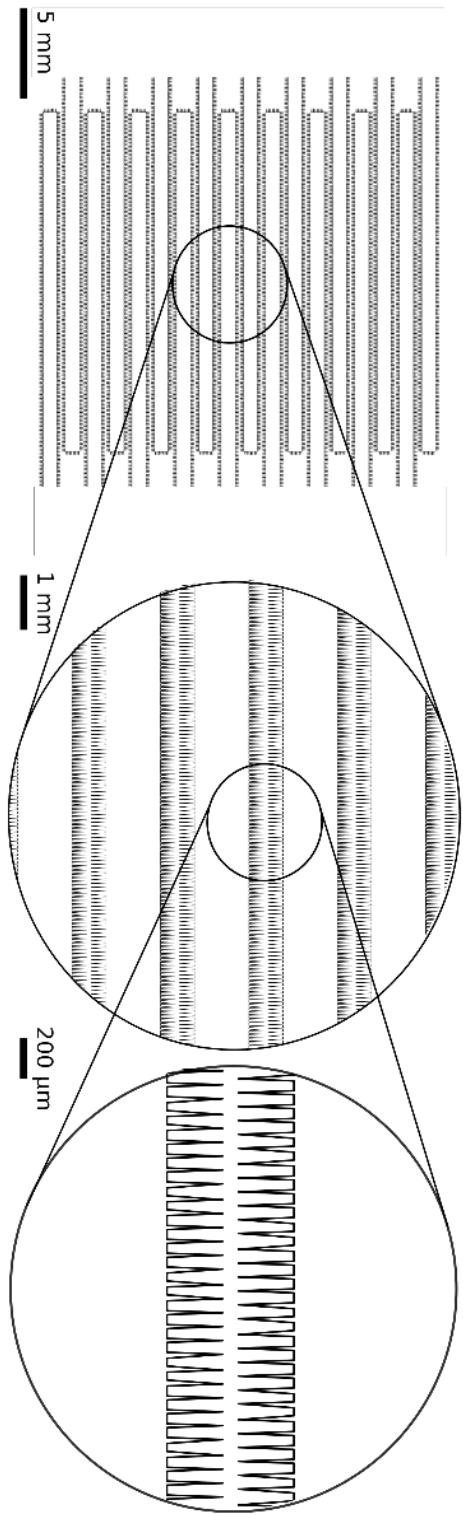

**Figure S1.** – Electrode design.

2) <u>Microparticle classification and reproducibility</u>

The microparticles were visually classified into trapped, oscillating, and attached states. The states were named based on the position and movement of the microparticles, but do not comprehensively describe all the motion patterns observed. Other forms of movement and organization of the microparticles, such as strings and collective motion, were observed when the microparticles were aggregated. We did not consider microparticles that were aggregated; therefore, we used a low concentration of microparticles to minimize particle-particle interaction. Only microparticles that were in the area of interest shown in Fig. S2 were considered. In some cases, microparticles would change from one state to the other; in this situation, the microparticle was not reclassified. Additionally, the behavior of the microparticles was highly dependent on the preparation method, that is, a sample that was only vortexed would have a different distribution of states compared to a sample that also was sonicated.

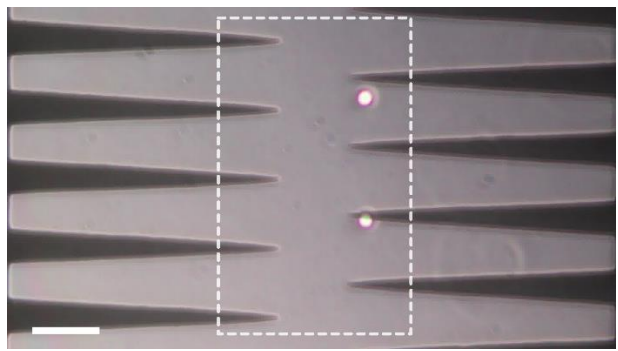

**Figure S1** – Area of interest where microparticles where classified. Scale bar = 70 μm.

3) <u>Measurement of background fluorescence</u>

To calculate the ratio between the fluorescence of microparticles and medium, we took fluorescent micrographs of 11 areas sized 832 μm 702 × 832 μm of a sample in a 0.2 mm thick polydimethylsiloxane PDMS gasket. The fluorescence density (total fluorescence a of region of interest divided by area) of the microparticles was measured. Then, image analysis software was used to exclude the microparticles and the fluorescence density of the background was measured. The background was mostly homogeneous, but surfactant aggregates with fluorescein could be observed at higher surfactant concentrations (Fig. S3).

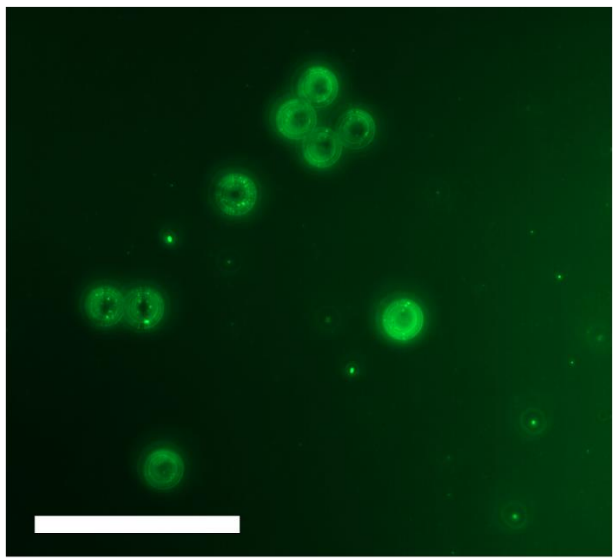

**Figure S3** – Particles covered with Span 80/fluorescein aggregates at the final Span 80 concentration of 5% (w/w). Scale bar = 200 μm.

## Supplementary Discussion

1) Langmuir adsorption model

The ratio of adsorbed and desorbed surfactant based on the surfactant concentration was fit with an equation derived from the Langmuir adsorption model[1]. The model assumes the surface of the microparticle is formed by discrete adsorption sites that the adsorbates, in this case the surfactant aggregates, can occupy (Fig. S4). The model considers that the adsorbates do not interact with each other and do not move, that the surface is homogeneous, and that the adsorption sites can only be occupied by one adsorbate at the time. We used this model as a first approximation with which to relate surfactant concentration and surfactant adsorbed. The model treats the adsorption as a reversible chemical reaction and gives the proportion of occupied sites in relation to the concentration of the adsorbent. In Eq. S1, $\theta_A$ corresponds to the fraction of the surface covered by the reverse micelles, $K_{ad}$ is the adsorption equilibrium constant, and $C_t$ is the surfactant concentration[138].

$$\theta_A = \frac{K_{ad} C_t}{1 + K_{ad} C_t} \quad (S1)$$

If it is assumed the maximum amount of charge the microparticle can have, $Q$, is proportional to the amount of surfactant on the surface of the microparticle, we obtain Eq. S2, as given below, where $q_{ar}$ is a constant that relates units of surfactants on the surface of the microparticle and charge:

$$Q(C_t) = q_{ar} \frac{K_{ad} C_t}{1 + K_{ad} C_t} \quad (S2)$$

To calculate the ratio of surfactant adsorbed on the microparticle and the surfactant in the medium, let $C_{max}$ be the maximum possible concentration of adsorbed surfactant and $C_{ad}$ the concentration of adsorbed surfactant. Therefore, the ratio of adsorbed to desorbed surfactant is $C_{ad}/(C_t - C_{ad})$. If we consider that the adsorbed surfactant is proportional to the occupied sites at the surface of the microparticle, we have Eq. S3, where $\alpha = C_{max}K_{ad}$ and $\beta = 1 - \alpha$ are constants.

$$C_{ad} = \frac{C_{max}K_{ad}C_t}{1 + K_{ad}C_t}$$

$$\frac{C_{ad}}{C_t - C_{ad}} = \frac{\frac{C_{max}K_{ad}C_t}{1 + K_{ad}C_t}}{C_t - \frac{C_{max}K_{ad}C_t}{1 + K_{ad}C_t}} = \frac{C_{max}K_{ad}}{1 + K_{ad}(C_t - C_{max})}$$

$$\frac{C_{ad}}{C_d} = \frac{\alpha}{\beta + K_{ad}C_t} = \frac{1}{\frac{\beta}{\alpha} + \frac{C_t}{C_m}} \qquad (S3)$$

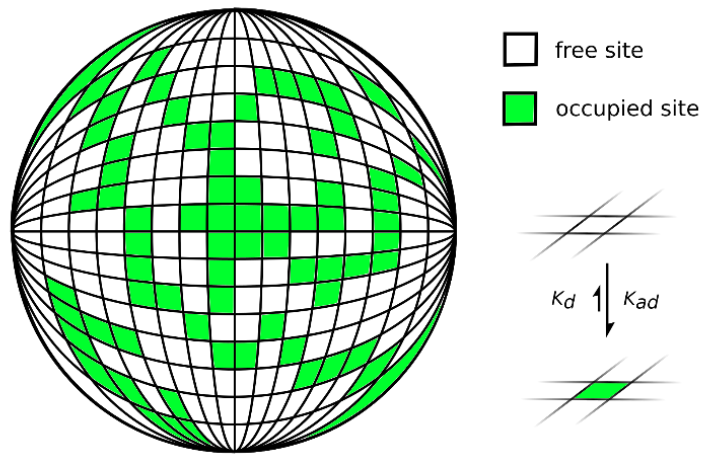

**Figure S4** – Langmuir adsorption model on the surface of a sphere.

2) Models of microparticle charging

Smith and McDonald[2] developed a theoretical model for charging of dust microparticles by ions. Based on their work, we establish an analogy to the charging of microparticles by charged reverse micelles, which provides an equation for the charging of microparticles mediated by surfactant. The original model described a system where dust microparticles are suspended in air that contains ions. When an electric field is applied, there is a flow of ions that charges the suspended microparticles[3,4]. In our system, the analogous insulating media is an apolar liquid instead of air, the ions are the charged reverse micelles, and the dust microparticles are the polystyrene microparticles. According to this model, the rate of charging is given by Eq. S4, where $q$ is the instantaneous charge of the microparticle, $Q$ is the saturation charge (or charge carrying capacity), $N_0$ is the charge density in the medium, $z$ is the electrical mobility of the charges and $\epsilon_m$ is the dielectric constant of the medium. $4\epsilon_m/(N_0 z Q)$ is a parameter of how fast the charging occurs; we denoted it $\tau_c$.

$$\frac{dq}{dt} = \frac{N_0 z Q}{4\epsilon_m}\left(1 - \frac{q}{Q}\right)^2$$

$$\frac{dq}{dt} = \frac{Q}{\tau_c}\left(1 - \frac{q}{Q}\right)^2$$

$$\frac{d(q/Q)}{d(t/\tau_c)} = \left(1 - \frac{q}{Q}\right)^2 \tag{S4}$$

However, Dascalescu et al.[5] noted that this charging rate is precise for microparticles charged by a current produced by a uniform electric field, which is not the case in our system. Taking that into consideration, we will assume that it remains valid as a qualitative description. In this model, the greater the charge the microparticle has, the slower it will charge $dq/dt \to 0$ as $q/Q \to 1$; therefore, it cannot have a charge greater than the charge carrying capacity $Q$ ($q \to Q$ as $t \to \infty$). To obtain the charge of a microparticle at time $t'$ that had charge $q_a$ when it touched the electrode at time $t'$, we can make a variable substitution, use the chain rule and integrate Eq. S4, resulting in the charging Eq. S5. For simplicity, we considered the initial time as zero. The effect of the parameters $Q, q(0), \tau_c$ on charging is illustrated on Fig. S5.

$$\int_{q_a/Q}^{q/Q} \frac{dx}{(1-x)^2} = \int_0^{t/\tau_i} dy$$

$$\frac{1}{1-x} \bigg|_{q_a/Q}^{q/Q} = \frac{t}{\tau_c}$$

$$q(t) = \frac{\frac{t}{\tau_c}(Q - q_a) + q_a}{1 + \frac{t}{\tau_c}\left(1 - \frac{q_a}{Q}\right)} \tag{S5}$$

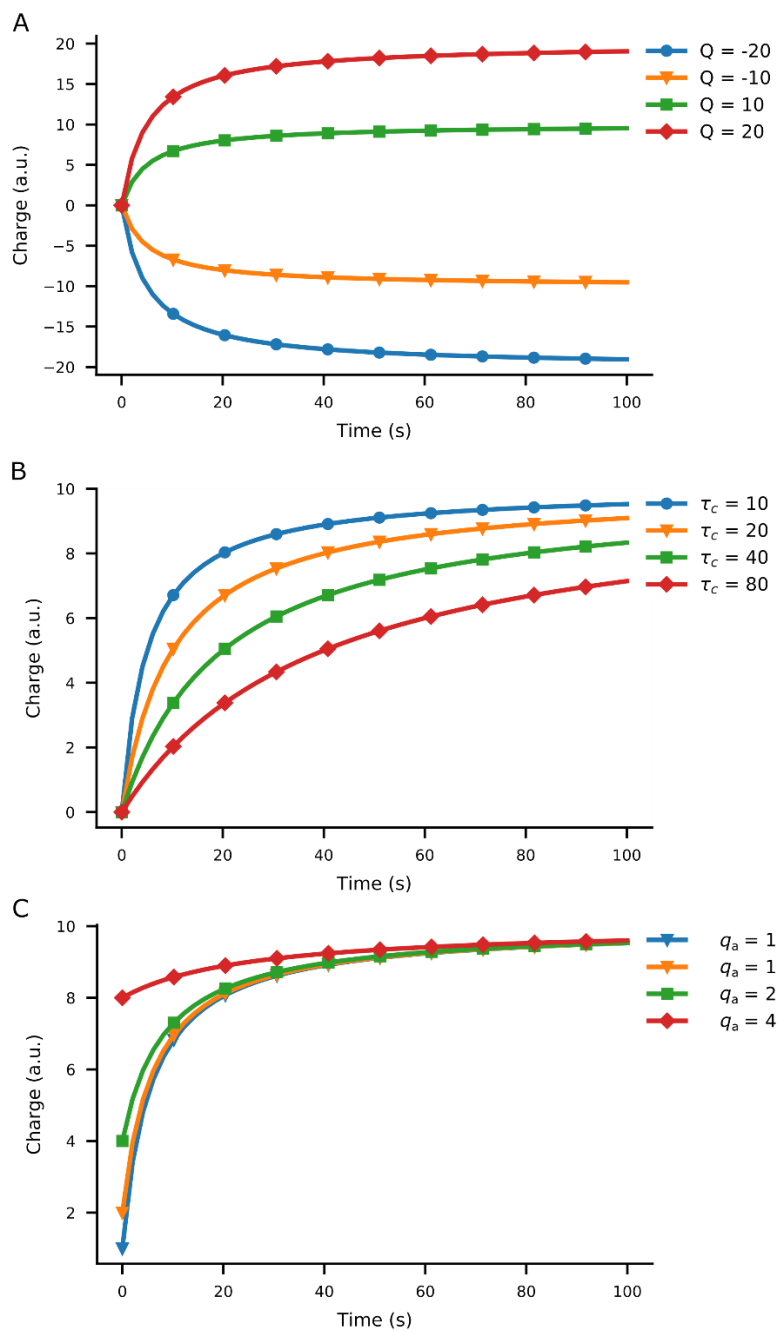

**Figure S5** – (A) Charging with different charging capacities $Q$, $q_a = 0$, $\tau_c = 5$. (B) Charging at different rates $\tau_c$, $Q = 10$, $q_a = 0$. (C) Charging at different initial charges $q_a$, $Q = 10$, and $\tau_c = 5$.

**Supplementary Notes**

1) <u>Choice of simulation parameters</u>

In our model, the microparticles are charged when they are in contact with the electrode, according to Eq. 4, and they are discharged due to charge relaxation according to Eq. 5. Although this mechanism is evidenced by the trajectory of the microparticles, it is not possible to obtain the charging and discharging parameters concomitantly from the analysis of the trajectory. To estimate the charge carrying capacity, we examined the order of the magnitude of the microparticle charges in apolar colloids with surfactants and the charge relaxation time of similar media. For instance, Espinosa et al.[6] studied the system of sub-micrometre PMMA microparticles in hexane with Span 85 and observed that they have charges in the order of tens of elementary charges. Microparticles of 20 μm would have a surface area 3 to 4 orders of magnitude larger, meaning that the average charge of a microparticle would be due to elementary changes of $10^4$ to $10^5$. Hsu et al.[7] observed that 780 nm PMMA microparticles in dodecane with dioctyl sulfosuccinate sodium salt (AOT) have $290 \pm 30$ elementary charges and proposes a model that constrains the maximum number of charges in the microparticles to $10^6$. For 20 μm diameter microparticles, that would mean microparticles with charges in the order $10^5$ elementary charges and limited to approximately $10^9$ charges. In our simulation, the charge carrying capacity varies between $\sim 10^6$ and $\sim 10^7$ elementary charges, depending on the surfactant concentration. Along these values, the charge relaxation time, $\tau_d$, was valued between $\sim 450$ s and $\sim 1$ s. For comparison, mica, which is used as an electric insulator, has a relaxation time of 51,000 s and corn oil, which is a weakly conductive oil, has a $\tau_d$ of 0.55 s[8].

2) Choice of initial conditions

In the experiments, the microparticles were uniformly distributed in the sample; therefore, we assumed random initial positions in the simulation. The initial velocity had a random orientation and a module following a Gaussian distribution of average zero and variance of 0.05 μm/s. For the initial charge of the microparticles, we considered the magnitude of the charges observed by Espinosa et al.[6] and Hsu et al.[7]. In our simulation, the microparticles started with a Gaussian distribution of charges with average charge zero and variance of $10^3$ elementary charges, which is equivalent to approximately 16 fC.

3) Microparticle classification in the simulation

The simulations lasted 200 s and the first and last 40 s were discarded. To classify the microparticles, we counted the number of times the microparticle left each electrode. If the microparticle did not leave the electrodes at any time and its position was between the electrodes (that is, it was not ejected from the simulation area), then it was considered trapped. If the microparticle left one electrode more than the other as defined by a tolerance then it was considered attached, and if it touched both electrodes for approximately the same number of times as defined by a tolerance parameter it was considered to be oscillating (see flowchart in Fig. S6). In our simulation, we used a tolerance value of 0.2.

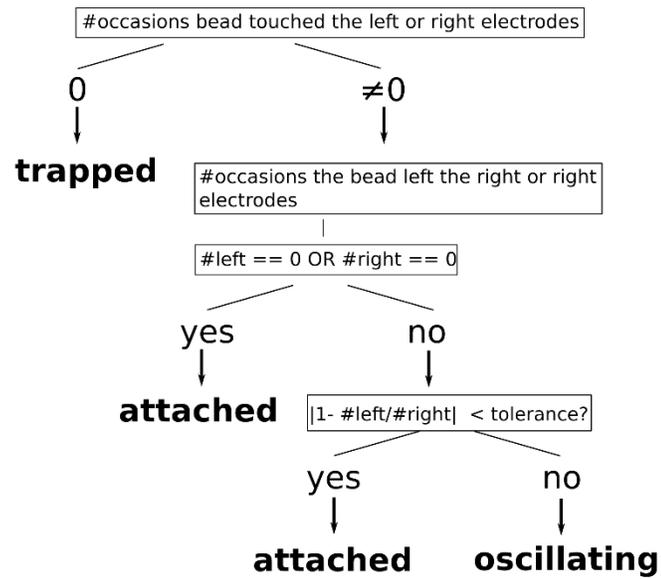

**Figure S6** – Algorithm for microparticle classification used in the simulation.

**Supplementary Videos**

Videos available at

https://takinouelab.github.io/MasukawaTakinoue2019/

**Movie S1**

Adsorption of Span 80 aggregates containing fluorescein on the microparticles does not hinder oscillation in the electric field produced by the sawtooth electrodes.

**Movie S2**

Particle trajectory simulation at a very low concentration of surfactant.

**Movie S3**

Particle trajectory simulation at a low concentration of surfactant.

**Movie S4**

Particle trajectory simulation at an intermediate concentration of surfactant.

**Movie S5**

Particle trajectory simulation at a high concentration of surfactant.

**Movie S5**

Particle trajectory simulation at a very high concentration of surfactant.

**Supplementary material references**